%% file: main.tex
\documentclass[11pt]{article}
\usepackage{lmodern}
\usepackage{dsfont}
\usepackage{environ}
\input{definitions.sty}

\input{macros.tex}

\begin{document}

\title{The Early Days of the Ethereum Blob Fee Market and Lessons Learnt}

\author{
 	    \textbf{Lioba Heimbach}\footnote{Work performed in part during an internship at a16z crypto.}
        \\
 		\small ETH Zurich \\
 		\small \emailhref{hlioba@ethz.ch}
        \and
 	    \textbf{Jason Milionis}$^*$\footnote{Jason's research was supported in part by NSF awards CNS-2212745, CCF-2212233, DMS-2134059, and CCF-1763970, by an Onassis Foundation Scholarship, and an A.G. Leventis educational grant.}
        \\
        \small Department of Computer Science \\
 		\small Columbia University \\
 		\small \emailhref{jm@cs.columbia.edu}
}
\date{Initial version: October 6, 2024 \\
      Current version: February 18, 2025
}
\maketitle

\thispagestyle{empty}

\begin{abstract}
Ethereum has adopted a rollup-centric roadmap to scale by making rollups (layer 2 scaling solutions) the primary method for handling transactions. The first significant step towards this goal was EIP-4844, which introduced blob transactions that are designed to meet the data availability needs of layer 2 protocols. This work constitutes the first rigorous and comprehensive empirical analysis of transaction- and mempool-level data since the institution of blobs on Ethereum on March 13, 2024. We perform a longitudinal study of the early days of the \emph{blob fee market} analyzing the landscape and the behaviors of its participants. We identify and measure the inefficiencies arising out of suboptimal block packing, showing that at times it has resulted in up to 70\% relative fee loss. We hone in and give further insight into two (congested) peak demand periods for blobs. Finally, we document a market design issue relating to subset bidding due to the inflexibility of the transaction structure on packing data as blobs and suggest possible ways to fix it. The latter market structure issue also applies more generally for any discrete objects included within transactions.
\end{abstract}

\hspace{3cm}
\begin{center}
    ~~\textbf{Keywords:} Ethereum, Layer 2, blobs, EIP-4844, rollup
\end{center}

\input{intro}

\input{analysis}

\input{discussion}

\section*{Acknowledgments}
We thank Edward Felten, Pranav Garimidi, Scott Duke Kominers, and Tim Roughgarden for helpful discussions on the subject matter of our paper.

\newpage   
\printbibliography
\newpage    
\appendix
\input{appendix}

\end{document}

%% file: macros.tex
\def\<{\langle}
\def\>{\rangle}

\newcommand{\emailhref}[1]{\href{mailto:#1}{\tt #1}}

%% file: intro.tex
\section{Introduction}

\newcommand{\nume}{num\'eraire\xspace}

\subsection{The Blob Fee Market}

Ethereum's journey towards enhanced scalability has been a central focus in its development roadmap~\parencite{buterin2024tweet}, driven by the need to address increasing network demand while preserving decentralization and security. As part of this effort, \textit{Ethereum Improvement Proposal (EIP)} 4844 introduced a novel transaction type that enables the use of \textit{blobs}---large, temporary data objects that are particularly critical for \textit{layer-2 (L2)} solutions. These blobs, by virtue of their enshrined short lifespans, are designed to offer a significant reduction in the cost of storing L2 data on-chain, thus aligning with Ethereum's broader goal of scaling through L2s.

The introduction of blobs under EIP-4844 also brought about a transformative shift in Ethereum's fee market by establishing a separate yet intertwined \emph{blob fee market}, a critical step in the paradigm of multi-dimensional resource pricing. Previously, there was a unified transaction fee priced after fixing the relative prices of all operations (including storage ones) in multiples of a single unit of account (the so-called gas unit), which was then multiplied by the fee given by the transaction originator per gas unit. Such a technique, however, inhibits a scenario where the relative prices of resources shift regimes through time, and would be particularly inflexible with very distinct resources constantly shifting demand like execution and (temporary) storage. With the advent of blob transactions, Ethereum also introduced a new, parallel, dynamically varying blob gas fee (the inner workings of which we elucidate in \Cref{subsec:blob_fees}) that varies separately from the (traditional) EIP-1559 \textit{execution gas fee}.

Our key focus is this \emph{blob fee market} and economic outcomes within this type of temporary storage, especially as they relate to rollups. While this fee market significantly reduced transaction fees for rollups, it also introduced complexities in terms of transaction structuring, optimal block packing, and economic incentives.

\subsection{Main Contributions}

This work delivers the first rigorous and comprehensive analysis of the Ethereum blob fee market during its initial months, empirically studying it and the behavior of its participants. Our extensive data collection and empirical analysis allows obtaining novel insights on the market internals of this temporary storage and contributes to the ongoing, active debate about the blob space on Ethereum.

Analyzing transaction- and mempool-level data we look into behavioral patterns of the market participants and their effects on-chain.
Two major events that spiked demand for blobs may provide a clearer picture of what a congested blob market would look like (especially in a world like the roadmap envisions with more active utilization of blobspace) and we hone in on these periods.

We show that a surprisingly large percentage of blocks are suboptimally packed with blob transactions, leading to up to 70\% fee losses for the builders who build these suboptimal blocks compared to the optimally built block.
We document that the market has been becoming slightly more efficient over time, but is still stunningly inefficient. We believe the reason for this to be the underinvestment in needed infrastructure due to the low amount of direct economic incentives.
Further than that, in \Cref{subsec:market}, we identify a structural market design flaw having to do with the rigid structure of transactions preventing the most efficient use of available blob space, that we term subset bidding.
We offer solutions and improvement suggestions for these inefficiencies of the market, as well as point out fruitful future avenues of research.
Importantly, we note that the market structure problem and solution we identify does \emph{not only} apply to blobs included in transactions, but \emph{any} potentially discrete ``object'' allocated in the same all-or-nothing way through the standard Ethereum transaction format.

\section{Background}

In the following, we introduce the relevant background regarding blobs on Ethereum (see \Cref{subsec:intro_blobs}), the fees paid by transactions including them (see \Cref{subsec:blob_fees}), and Ethereum block building (see \Cref{subsec:block_building}).

\subsection{Blobs}
\label{subsec:intro_blobs}
To tackle Ethereum scalability issues, EIP-4844 introduced a new type of transactions---referred to as \textit{type-3 transactions} or \textit{blob transactions}---which allow the sender to submit blobs of data. L2s primarily use these blobs for transaction settlement on the L1. Blobs only persist for a short period (around 18 days) on the network, long enough for them to be retrieved but short enough to prevent excessive long-term usage of storage. This temporary nature allows blobs to be priced lower than calldata,\footnote{Before blobs were introduced, L2s used Ethereum calldata for settlement of L2 data on the L1.} which must be permanently stored on the blockchain and contribute to state growth. Up to six blobs may be included on each block, and a type-3 transaction can have between one and six blobs. There is space for 128kb of data in each blob and a transaction always pays for the entire blob even if it only uses a partial portion of the allowable storage space it provides.

Similarly to all other types of transactions, type-3 transactions that are publicly broadcast are stored in Ethereum's execution layer network \textit{mempool}, i.e., the public waiting area for transactions. Importantly, the blob data itself is not propagated in the execution layer network. Instead, blobs are propagated in the consensus layer network. Thus, type-3 transactions propagated through the execution layer network and included on-chain will only contain a reference to the blob.

\subsection{Blob Fees}\label{subsec:blob_fees}
Type-3 transactions simultaneously pay gas fees in: (1) the ``normal'' Ethereum gas market, and (2) the blob gas market.

Similar to other types of transactions, type-3 transactions can include calldata, transfer Ether, or interact with a smart contract. The fees paid by a type-3 transaction $tx$ included in block $n$ in the ``normal'' gas market, which we will refer to as the EIP-1159 gas market throughout, are as follows.

$$ \textsc{fee}_{1559} (tx,n) = \textsc{gas}(tx) \cdot (\textsc{base\_fee}(n) + \textsc{priority\_fee}(tx)),$$
where $\textsc{gas}(tx)$ is the transaction gas usage. Further, $\textsc{base\_fee}(n)$ is the block's base fee charged per unit of gas: it represents the minimum fee transactions must pay in the block and automatically updates based on past block gas usage~\parencite{ethereum2024eip1559}.  Finally, $ \textsc{priority\_fee}(tx)$ is the transaction's priority fee charged per unit of gas. Importantly, the part of the fee paid by transaction $tx$ associated with the base fee (i.e., $\textsc{gas}(tx) \cdot \textsc{base\_fee}(n)$) is burned. Only the remaining fees ($\textsc{gas}(tx) \cdot \textsc{priority\_fee}(tx)$ are received by the block's fee recipient. 

Additionally, type-3 transactions also pay fees per blob included in the blob gas market, which we will refer to as the EIP-4844 gas market throughout. To be precise, the fees a transaction $tx$ included in block $n$ pays in the EIP-4844 gas market are
$$ \textsc{fee}_{4844} (tx,n) = \textsc{num\_blobs}(tx) \cdot \textsc{blob\_base\_fee}(n),$$
where $\textsc{num\_blobs}(tx) $ is the number of blobs included by the transaction and $ \textsc{blob\_base\_fee}(n)$ is the blob base fee in block $n$ which is charged per blob gas. The blob base fee for block $n$ is derived as follows
$$\textsc{blob\_base\_fee}(n)=\textsc{min\_fee}\cdot \exp{\left(\frac{\textsc{total\_excess\_gas}(n-1)}{\textsc{update\_fraction}}\right)},$$
where $\textsc{total\_excess\_gas}(n-1)$ is the total blob gas used in excess of the target before the current block. Note that while there is space for six blobs per block, the target is three. Thus, whenever more than three blobs are included in a block the excess increases and decreases when less than three are included. Additionally, $\textsc{min\_fee}$ is a constant currently set to 1 wei, and $\textsc{update\_fraction}$ is a constant set such that the maximum increase in the blob base fee per block is 12.5\%~\parencite{ethereum2024eip4844}. Importantly, all fees associated with the EIP-4844 gas market are burned, i.e., the block's fee recipient receives no fees for blobs included.

\subsection{Block Building}\label{subsec:block_building}
The vast majority of blocks ($\approx$90\%) in Ethereum are built through a scheme called \textit{Proposer-Builder Separation (PBS)}~\parencite{wahrstatter2024}. With PBS the validator chosen as the block proposer is only responsible for proposing the block, while specialized \textit{builders} are responsible for building the blocks. The idea is that these specialized builders are better at building high-value blocks, i.e., blocks with significant fee revenue. Additionally, these specialized builders likely have access to value private order flow, i.e., transactions that are not broadcast to the public mempool. Note that in the current implementation of the scheme, validators and builders communicate with each other through a relay: a party trusted by the two~\parencite{flashbots2024}.

\section{Related Work}
\label{subsec:litrev}

\paragraph{EIP-1559 fee market} The first major shift in the Ethereum transaction fee mechanism was the deployment of EIP-1559 in 2021. Before deployment, \textcite{roughgarden2020transaction} presented a game-theoretic analysis of the mechanism, while \textcite{leonardos2021dynamical,ferreira2021dynamic} conducted studies focusing on the dynamic update rule of the base fee. On the empirical side, multiple studies \parencite{reijsbergen2021transaction,liu2022empirical,leonardos2023optimality} demonstrate that the introduction of EIP-1559 made gas fees more predictable despite the short-term oscillation in block size. In addition, for farsighted validators, it can be rational to attack the mechanism leading to greater unpredictability of fees~\parencite{hougaard2023farsighted,azouvi2023base}. On the other hand, our work focuses on the blob fee market, which was introduced through EIP-4844.

\paragraph{Multidimensional fee markets} A recent line of literature explores multidimensional blob markets. \textcite{diamandis2023designing} propose an efficient pricing mechanism for multiple resources, while \textcite{angeris2024multidimensional} show that these multidimensional blockchain fee markets are essentially optimal. In contrast, we empirically explore the blob fee market, which added a second dimension to Ethereum's fee market.

\paragraph{Blob fee market} \textcite{crapis2023eip} conducted an economic analysis of the blob market at its launch, examining whether rollups would prefer posting data in blobs or calldata (the original market). Their study concludes that large rollups with high transaction rates would opt for blobs, while those with lower rates would favor the original market. Our empirical findings, however, show that even when the blob market is congested and more expensive than calldata, rollups do not revert to the original market.

\textcite{soltani2024delay} analyze the delays of blob transactions in a \emph{time-average} fashion to conclude that average delays are lower if all transactions only carry a small number of blobs; this is a consequence of mainly standard limited-resource congestion. Contrary to that, one of our multifaceted contributions deals with examining the delays as a consequence of inefficiencies in block packing, where we study how this has changed \emph{throughout time} and its sensitivity, especially in demand-congested periods.

In the period before the introduction of blobs, \textcite{messias2024writing} performed a comprehensive transaction analysis focusing on inscriptions, which triggered the first significant demand for blob space on Ethereum. Their work, however, does not delve into the internal workings of the blob market, which is the central focus of our study.

A recent research post performs an empirical evaluation of the blob market, focusing on the blob base fees paid and the consequences of increasing the parameter of the minimum base fee~\parencite{dataalways2024minimumblobfees}. Concurrent work by \textcite{lee2024180} explores the possibility of rollups sharing blob space among each other (a practice called \textit{blob sharing}) and simulates the impact using the first 180 days of blobs on Ethereum. Their goal is to minimize the costs for rollups. In contrast, our work focuses on the \emph{blob fee market} as a whole and its participants (with an emphasis on \emph{block builders}) while also identifying and documenting market design issues, offering potential solutions to those.
Finally, likewise, auction considerations have been previously connected to market mechanisms, including in decentralized finance infrastructure \parencite{myersonian_mm,complexity,nftauctions}.

%% file: analysis.tex
\section{Data Collection}\label{sec:data}
To study the Ethereum blob market, we gather three different types of data: Ethereum blockchain, PBS, and Ethereum mempool data. Our data collection spans the range from block 19,426,589 -- the block marking the Dencun hardfork that introduced blobs on March 13, 2024 -- to block 20,866,918, capturing the final block as of September 30, 2024. Consequently, this dataset encompasses the entire timeline of the blob market's activity up until the end of September 2024. %

\paragraph{Ethereum Blockchain}
We run a Reth execution layer node~\parencite{reth2024} and a Lighthouse consensus layer node~\parencite{sigmaprime2024lighthouse} to gather Ethereum blockchain data. We parse the blockchain for blob transactions, recording relevant details such as the number of blobs submitted and the fees paid. Type-3 transactions pay fees for both regular Ethereum gas in the EIP-1559 market (base and priority fees) and blob gas in the EIP-4844 market (blob base fee). Additionally, transactions may tip block fee recipients via direct transfers.

\paragraph{PBS}
We collect data from eight relays that were active during the study period: Aestus, Agnostic, bloXroute (Max Profit), bloXroute (Regulated), Eden, Flashbots, Manifold, and UltraSound. These relays provide public APIs that allow access to the blocks they deliver to proposers. We use the PBS data to obtain information about the block builder. The reason we use this builder-specific information is to investigate whether specific block builders have different strategies on (optimal) transaction choice with blob transactions.

\paragraph{Ethereum Mempool}
Finally, we gather Ethereum mempool data from the Mempool Dumpster project~\parencite{flashbots2024mempool}. The mempool data allows us to observe the blob mempool, i.e., the blobs waiting for block inclusion in the public P2P network, so that we can then analyze the efficiency of blob inclusion for each builder and the waiting time for blob transactions to be posted on-chain.
It is important to note that there is minimal private submission of blob transactions throughout the entire period, and negligible during the congested periods that we focus on (for methodological details, we refer the interested reader to Appendix~\ref{app:private}).

\section{Blob Adoption}
\begin{figure}[t!]\vspace{-6pt}
    \centering
    \begin{subfigure}[t]{0.48\columnwidth}
    \centering
        \includegraphics[scale=1.15]{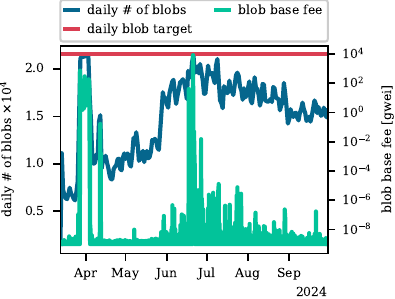}
    \caption{Daily number of blobs (green line) and \textbf{daily} blob target (red line) on the left y-axis. We further plot the blob base fee (blue line) on the right y-axis.}
    \label{fig:adoption2}
    \end{subfigure}
    \hfill
    \begin{subfigure}[t]{0.49\columnwidth}
    \centering
        \includegraphics[scale=1.15]{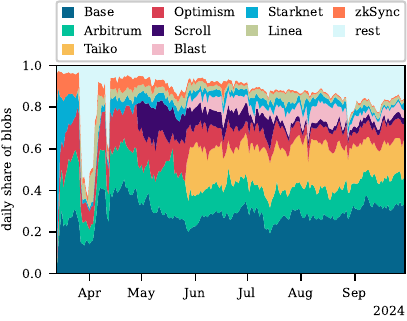}
    \caption{Daily share of blobs submitted by L2s. The biggest nine blob submitters are shown separately and the rest are grouped together. }
    \label{fig:shareOfBlobs}
    \end{subfigure}    \vspace{-6pt}
    \caption{Blob usage by L2s. Figure~\ref{fig:adoption2} shows the overall demand, while Figure~\ref{fig:shareOfBlobs} visualizes the individual blob usage of the biggest L2s.}\label{fig:adoption}
\end{figure}

We commence our analysis by analyzing the adoption of blobs. \Cref{fig:adoption2} visualizes the daily number of blobs included on Ethereum along with the daily target (left y-axis). While we notice a general increase in the daily number of blobs, the number of blobs has not sustainably reached the target. Instead, they have only approached the target on two occasions: (1) the Blobscription heavy interest period (``craze'') starting at the end of March 2024~\parencite{nijkerk2024ethereum} and (2) the LayerZero airdrop on Arbitrum on June 20, 2024~\parencite{stevens2024arbitrum}. During these demand increases, we also observe very significant spikes in the blob base fee (see right y-axis in Figure~\ref{fig:adoption2}). Astonishingly during the second incident blob fees increase by nearly 15 orders of magnitude from $1$~wei (the minimum blob base fee) to $10^{15}$~wei. However, while there was little variance (except for the individual incidents) in the first two months of the blob market, the blob base fee variance has risen from June 2024 along with the general increase in blob usage. At the same time, this variance has come down slightly in the past months as demand for blobs has also decreased again. Given that there is no sustained demand for blobs at their target number, the blob base fees have generally been meager, as market economics of demand and supply would dictate.

In \Cref{fig:shareOfBlobs} we visualize the proportion of blobs posted by individual senders on a daily basis. The nine largest L2s, in terms of the total number of blobs submitted, are shown separately and ordered by size while the rest are grouped together. We start by noting that, generally, these nine L2s are responsible for more than 80\% of blob usage. This figure decreases slowly during our measurement period, indicating that blob demand has extended to a wider group of projects. Additionally, blob demand extended beyond the biggest L2s surrounding the Blobscription craze starting at the end of March 2024. In regards to the biggest L2s, we notice that Base consistently uses more than one-fifth of blobs daily, with Arbitrum being the second biggest blob user. After that, the picture becomes more fragmented with some L2s only starting to post blob a couple of months into our data collection window (e.g., Taiko) or losing market share with time (e.g., zkSync). Overall, the market appears to grow less concentrated over time.

\subsection{Fees}
\begin{figure}[t]\vspace{-6pt}
\centering

\begin{minipage}[t]{.48\linewidth}
    \centering
    \includegraphics[scale=1.2]{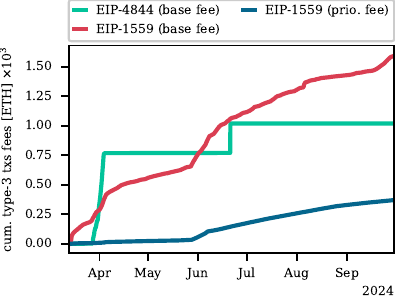}\vspace{-6pt}
    \caption{Cumulative fees paid by blob transactions over time. We separate the EIP-4844 fee market (base fee) from the EIP-1559 fee market (base and priority fee).}
    \label{fig:fees}
\end{minipage}
\hfill
\begin{minipage}[t]{.48\linewidth}
    \centering
     \includegraphics[scale=1.2]{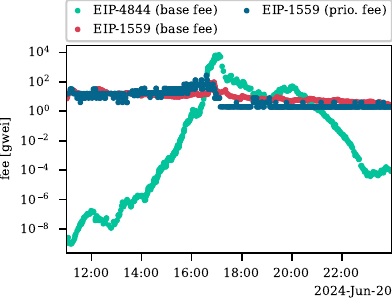}\vspace{-6pt}
    \caption{Development of various fee components during the spike in blob demand caused by LayerZero airdrop on 20 June 2024.}
    \label{fig:feesLayer0}
\end{minipage}
\end{figure}

Recall, that since the introduction of blobs on Ethereum, the blob base fee has been very low outside of our two outlined major events that spiked demand. In the following, we analyze how the blob base fees paid by type-3 transactions compare to the fees they pay in the EIP-1559 fee market. \Cref{fig:fees} visualizes the cumulative fees paid by type-3 transactions. The cumulative EIP-4844 base fee paid increased abruptly on two occasions: Blobscriptions and the LayerZero airdrop. Outside of these events, the cumulative EIP-4844 base fee appears almost constant. The two EIP-1559 market fee components (base and priority fee) increase much more steadily. For both, we notice an increase in growth at the beginning of June 2024, which corresponds to a general increase in blob demand (see \Cref{fig:adoption2}). Overall, blob transactions have paid 1,020~ETH in the EIP-4844 base fee, 1,602~ETH in the EIP-1559 base fee, and 372~ETH in the EIP-1559 priority fee. While the 2,993~ETH (approximately US\$~8~M) paid in total fees by type-3 transaction initially appears large, this only amounts to 1\% of the fees paid on Ethereum in the same period. Furthermore, only the 372~ETH in priority fees are not burned given that no type-3 transaction has included a coinbase transfer. Thus, these 372~ETH are the sole financial incentive for block builders to include type-3 transactions (i.e., approx US\$~0.7 per block at the current ETH price). In contrast, the average fee revenue from a block is around US\$~200. These small financial incentives builders have to include type-3 transactions that hint at some of the present issues in the blob market (see \Cref{sec:issues}).

We further analyze how the various fee components react to a sudden spike in demand and plot the various fee components from 11:00 to 23:00 on 24 June 2024 in Figure~\ref{fig:feesLayer0}. For each blob transaction, we draw a dot for each of the three components. First note that while the EIP-4844 base fee increases by around 15 orders of magnitude within six hours, there is comparatively little movement in the EIP-1559 fee market. Blob transactions seem to increase their priority fees slightly (by around one order of magnitude), and the EIP-1559 base fee is largely unaffected given that blob transactions are only a small proportion of the overall transaction demand (i.e., blob transactions make up around 0.5\%). We further highlight that the EIP-1559 fees recovered a lot quicker, while it took some time for the EIP-4844 to come down again. In addition to taking a long time to recover, the base fee also took several hours to increase. This slow price discovery is a consequence of the blob base fee rule update rule (see \Cref{subsec:blob_fees}) taking more than ten blocks to increase or decrease by an order of magnitude. 

As a consequence of the slow price discovery, one would expect the rollups to compete in a first price auction~\parencite{buterin2024multidimensional,dataalways2024minimumblobfees}. However, we only mildly observe this behavior as the priority fee only changes by around one order of magnitude, which does not reflect the 15 orders of magnitude jump of the base fee. Additionally, to the best of our knowledge, our data indicates that none of the rollups moved to the original market and used calldata even though it would have been cheaper to do so---a sign that they were not prepared for such an extreme congestion event.

\subsection{Behavior of L2s}
\begin{figure}[t]\vspace{-6pt}
    \centering
    \includegraphics[scale=1.2]{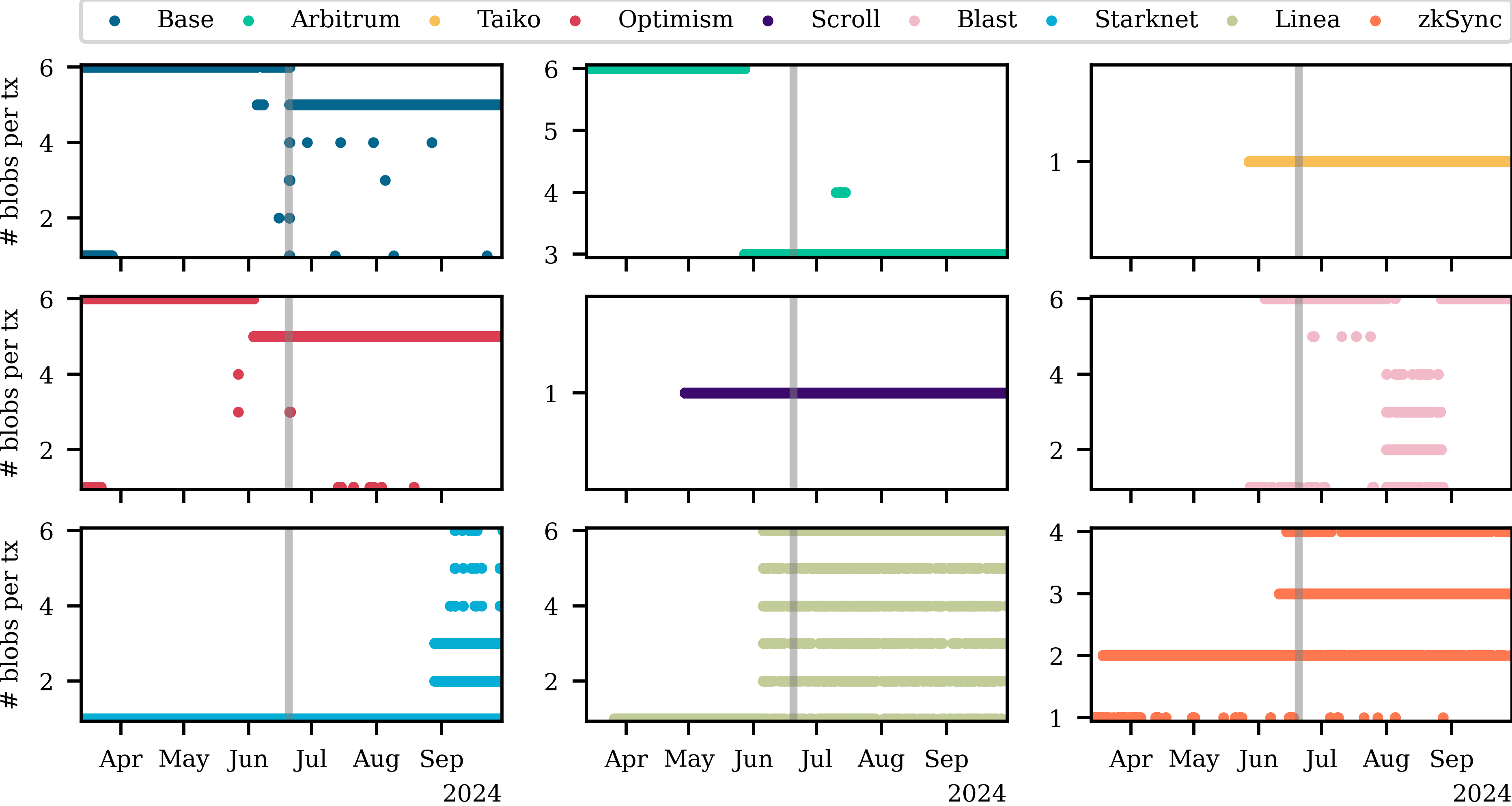}\vspace{-6pt}
    \caption{Number of blobs submitted by the biggest nine L2s per type-3 transaction over time, each dot represents one type-3 transaction submitted by the respective L2. The grey vertical line indicates the spike in demand on 20 June 2024 related to the LayerZero airdrop. }
    \label{fig:numBlobs}
\end{figure}

In the following, we analyze the behavior of L2s when submitting type-3 transactions, focusing on the number of blobs per transaction and how this changes over time. Figure~\ref{fig:numBlobs} shows different strategies by L2s: Taiko and Scroll consistently submit one blob, while others, like Blast, Starknet, Linea, and zkSync, vary their blob counts. These four potentially adjust to demand on their own networks or demand for blobs. Finally, large L2s (e.g., Base, Arbitrum, Optimism) tend to use a fixed number of blobs but adjust their strategy over time. Potentially reacting to long-term changes in demand, and on specific occasions (i.e., the LayerZero airdrop indicated by the vertical gray line). One change we observed among most type-3 transaction senders is that starting around June 2024, many of them either diversified how they submit type-3 transactions or adjusted their strategy. This is likely a result of the long-term increase in demand picking up in June 2024. For example, Base and Optimism moved from submitting six blobs per transaction, i.e., taking up the entire blob space in a block, to only submitting five per transaction, i.e., leaving space for one additional blob. Arbitrum similarly adjusted from submitting six blobs per transaction to three. These adjustments could also be related to delays experienced as a result of inefficiencies in the blob market which we will discuss in detail in Section~\ref{sec:issues}.

\begin{figure}[t]\vspace{-6pt}
    \centering
    \begin{subfigure}[t]{1\columnwidth}
        \includegraphics[scale=1.2,right]{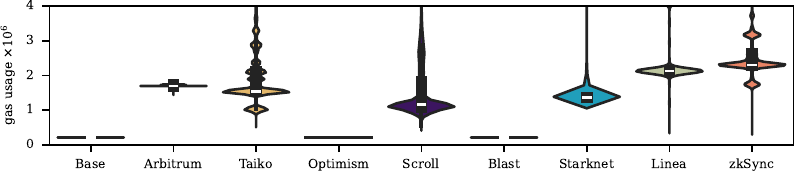}
    \caption{gas usage}
    \label{fig:gas_used}
    \end{subfigure}    
    \begin{subfigure}[t]{1\columnwidth}
        \includegraphics[scale=1.2,right]{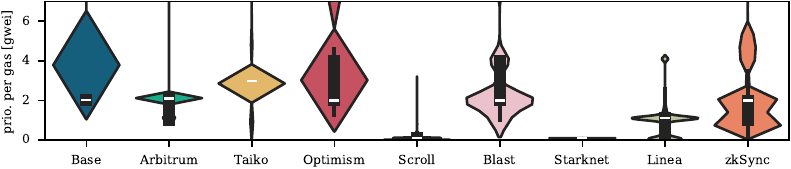}
    \caption{priority fee}
    \label{fig:prio}
    \end{subfigure}
    \begin{subfigure}[t]{1\columnwidth}
        \includegraphics[scale=1.2,right]{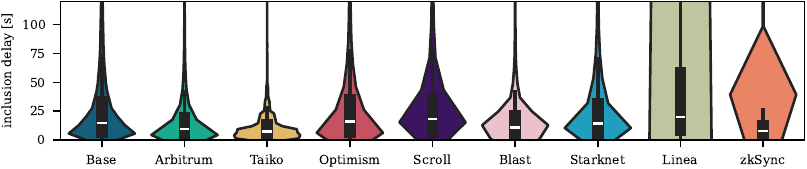}
    \caption{inclusion delay}
    \label{fig:inclusionDelayMs}
    \end{subfigure}\vspace{-6pt}
    \caption{Distribution of priority fee (see Figure~\ref{fig:prio}), gas usage (see Figure~\ref{fig:gas_used}) and inclusion delay  (see Figure~\ref{fig:inclusionDelayMs}) of type-3 transactions submitted by the nine biggest L2s.}\label{fig:gas}    
\end{figure}

Next, we consider the priority fees and gas usage of type-3 transactions in the EIP-1559 market (see \Cref{fig:gas}). Recall, that type-3 transactions, in addition to paying for the blobs they include, also pay fees in the EIP-1559 market for any gas they use. In \Cref{fig:gas_used}, we make a violin plot of the gas usage of the biggest L2s for their type-3 transactions. Note that a violin plot combines a boxplot, which displays the lower quartile, median, and upper quartile, with a kernel density plot to represent the distribution of values. Base, Optimism, and Blast consistently use the minimum amount of gas possible on Ethereum: 21,000~\parencite{wood2014yellowpaper}. Note that these are all optimistic rollups, i.e., they only post data. The two further optimistic rollups (i.e., Arbitrum and Taiko) have significantly higher gas usage but that of Arbitrum is still extremely consistent. For the four zk rollups (i.e., Scroll, Starknet, Linea, zkSync), we generally observe a higher gas usage with higher variance. Note that this is expected as zk rollups execute validity proofs that verify the execution of their transactions. 

Regarding priority fees, we also observe different patterns between the biggest type-3 transaction senders (see Figure~\ref{fig:prio}). Scroll and Starknet consistently have extremely low priority fees with averages of 0.07 and 0.1~gwei respectively. Note that all type-3 transactions from these two senders are below 1~gwei which is the minimum priority fee accepted by the geth builder~\parencite{szilagyi2024tweet5} -- the biggest execution layer client~\parencite{etheralpha2024}. The remaining senders all have similar priority fees between 1.5~gwei and 3.5~gwei but different distributions. Interestingly, type-3 transactions only have 1.9~gwei average priority fee, and around 20\% have a priority fee below 1~gwei. Further, this figure is also lower than the average priority fee of all transactions (i.e., 3.2~gwei during the same period).

We further note that the L2s that posted type-3 transactions with different numbers of blobs consistently for an extended period often do not increase the priority fee depending on the number of blobs they include. As we show in Appendix~\ref{app:prio}, there is no clear positive correlation between the number of blobs and the priority fees for all of them. 

Finally, we visualize the inclusion delay in seconds for each of the biggest nine blob posters in Figure~\ref{fig:inclusionDelayMs}. Taiko type-3 transactions wait for the shortest time on average with ten seconds, two seconds less than the time between two blocks. On, the other hand, Linea type-3 transactions wait for 100 seconds on average, i.e., more than eight blocks. Interestingly, Scroll and Starknet which have the lowest priority fees, do not wait for block inclusion significantly longer than many of the other senders that pay much higher priority fees.

\section{Inefficiencies in the Blob Market}\label{sec:issues}
In the previous sections, we hinted at L2s potentially adjusting the number of blobs they include per transaction to respond to the delay they possibly experience as a result of inefficiencies in the blob market. We look into these issues in this section.

\subsection{Block Packing}
Recall that there is space for at most six blobs in a block and that each transaction can include between one and six blobs. Solving the packing problem optimally is equivalent to solving knapsack which is NP-hard to optimally solve. On the other hand, a naive, greedy implementation would sort type-3 transactions by priority fee and select transactions as long as there is still space. To see why this is not optimal consider the following example where each transaction is a tuple $(\textsc{gas\_usage},$$\textsc{num\_blobs},$$ \textsc{priority\_fee})$: there are three type-3 transactions represented by $(1,5,2)$, $(1,3,1.99)$ and $(1,3,1.99)$. The naive algorithm would choose the first transaction but would have no space to include the other two while the optimal algorithm (assuming there is enough gas) would pick the last two transactions and receive nearly double the fees (i.e., 3.98 instead of 2).

Despite the worst-case infeasibility of solving knapsack (an NP-hard problem), since there is only space for six blobs, one can actually \emph{easily} solve the blob packing problem optimally and reasonably quickly. We do these calculations to investigate whether blobs in blocks were optimally packed with respect to the builder's fee income. In case they were not, we quantify the suboptimality, which at times reaches high double-digit percentages (up to 70\% suboptimal).
For this, we use the mempool data to determine which type-3 transactions were waiting in the mempool. Importantly, we consider only type-3 transactions that were included in a later block and were first observed in the mempool at least four seconds before the block's expected time, but no more than 120 seconds prior. This ensures the block builder had sufficient time to see the type-3 transaction and that it was unlikely to be cleared from the builder’s local mempool. We further note that most transactions are submitted to the public mempool except for type-3 transactions from Taiko being submitted privately starting from September 2024 (see Section~\ref{app:private}). Finally, we exclude blocks without blobs from this analysis as not including blobs by the block builder might be a choice. In summary, whenever we observe suboptimal blob packing we can be fairly certain that it was suboptimal but the actual suboptimality may have been slightly \emph{higher}.

\begin{figure}[t!]\vspace{-6pt}
    \centering
    \begin{subfigure}[t]{0.49\columnwidth}
    \centering
        \includegraphics[scale=1.2]{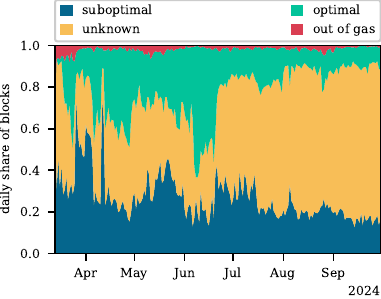}
    \caption{PBS}
    \label{fig:optimal}
    \end{subfigure}\hfill
    \begin{subfigure}[t]{0.49\columnwidth}
    \centering
        \includegraphics[scale=1.2]{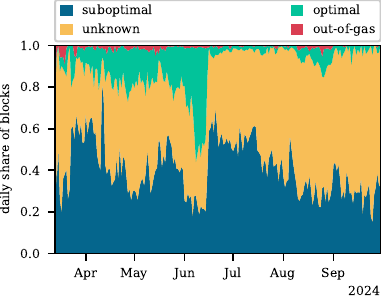}
    \caption{non-PBS}
    \label{fig:vanilla}
    \end{subfigure}\vspace{-6pt}    
    \caption{Daily breakdown of blocks by blob packing: suboptimal, optimal, matching naive and optimal algorithms (``unknown''), or out of gas with optimal packing. Figure~\ref{fig:optimal} shows the breakdown for PBS blocks and Figure~\ref{fig:vanilla} for non-PBS blocks.}\label{fig:optimalpacking}
\end{figure}

We start by analyzing what proportion of blocks are packed optimally, suboptimally, whether the naive algorithm achieves the same result as the optimal one (i.e., what we call ``unknown''), or whether the block would exceed the maximum allowable gas in an Ethereum block with the optimal blob packing\footnote{Therefore, in this ``out of gas'' case, the builder has a clear reason not to include this particular transaction combination yielding the optimal blob packing, as it would lead other potentially profitable bundles to not be able to be included in the built block.} (i.e., what we call ``out of gas''). We do the analysis both for block build through PBS (see Figure~\ref{fig:optimal}) as well as non-PBS blocks (see Figure~\ref{fig:vanilla}). We start by noting that on average there are 1.9 blobs per block for PBS blocks and 2.2 blobs per block for non-PBS blocks. Further, PBS blocks are built by specialized builders, while non-PBS blocks are less likely to be. 

Thus, one would expect blobs in PBS blocks to be packed optimally with respect to the block builder's fee income. In \Cref{fig:optimal}, we notice that this is stunningly not the case. Instead, we can only be sure that 19\% of blocks PBS had an optimal blob packing, while 44.6\% had a suboptimal packing. Furthermore, only a very minor proportion of blocks would exceed the gas limit with optimal blob packing, running into an execution layer restriction that would prevent them from including the income-optimal number of blobs. In general, we notice that over time the proportion of suboptimal blocks has decreased, while the proportion of blocks where it is unknown whether the blob packing algorithm is optimal is increasing. In part, this could be due to type-3 transactions including fewer blobs on average over time (see Figure~\ref{fig:numBlobs}). Additionally, there may be several reasons why specialized builders pack blobs suboptimally.
For one, the investment into an optimal blob building might not be worth it given the low fee revenue (see Figure~\ref{fig:fees}).
Another possibility is related to a race condition (time taking for blocks with more data, i.e., blobs, to propagate across clients) and timing games being played by builders submitting late blocks into the slot to the PBS system. Builders might submit blocks with a varying number of blobs to relays, hence for latency reasons, sometimes blocks with no blobs win the PBS auction and get published instead of their blob-including counterparts.

Figure~\ref{fig:vanilla} visualizes the same data for non-PBS blobs. We notice that while there are more suboptimal blocks (i.e., 47\% on average) and less optimal blocks (i.e., 9\% on average), the difference between PBS and non-PBS blocks is not significant in terms of blob packing. This contrasts with the comparison of the two in terms of overall fee revenue, MEV, etc.~\parencite{heimbach2023ethereum}. 

\begin{figure}[t!]\vspace{-6pt}
    \centering
    \begin{subfigure}[t]{0.49\columnwidth}
    \centering
        \includegraphics[scale=1.2]{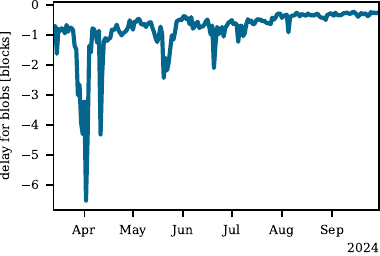}
    \caption{type-3 transaction delay}
    \label{fig:delay}
    \end{subfigure}\hfill
    \begin{subfigure}[t]{0.49\columnwidth}
    \centering
        \includegraphics[scale=1.2]{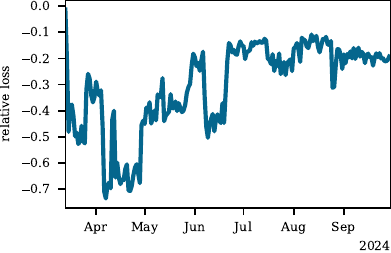}
    \caption{relative blob priority fee loss}
    \label{fig:loss}
    \end{subfigure}\vspace{-6pt}
    \caption{Delay (see Figure~\ref{fig:delay}) and relative blob priority fee loss (see Figure~\ref{fig:loss} as a result of suboptimal blob packing. }\label{fig:issues}
\end{figure}

Finally, notice that the number of suboptimal blocks peaked at the end of March 2024 during the Blobscriptions craze -- the first peak in blob demand. In the following, we investigate the effects of these suboptimal blob packings in detail (see Figure~\ref{fig:issues}). We start by analyzing the delay in Figure~\ref{fig:delay} and notice that in general, the blobs do not have to wait more than two blocks as a result of suboptimal blob packing. The one big exception is the Blobscriptions craze starting at the end of March 2024, during this period type-3 transactions were delayed by an average of more than eight blocks as a result of suboptimal packings. Regarding loss in fee revenue for the builders of blocks with suboptimal blob packings, we observe a slightly different packing. In general, with time the relative loss decreases. Further, the relative loss is highest during the Blobscriptions craze reaching a relative loss of 70\%. There is a further increase in the relative loss during the increase in blob demand starting at the beginning of June which peaked during the LayerZero airdrop at the end of June.

\subsection{Market Design Issue: Subset Bidding}
\label{subsec:market}

Beyond the above blob packing issues, a different but related---this time fundamental to market design---problem has to do with the transaction format of blob inclusion.
In short, recall that a blob (also known as type-3) transaction on Ethereum contains a number of blobs, between 1 and 6 blobs. Only 6 blobs may be included per block, regardless of the number of transactions.
The key part is that, if a blob transaction is included, then \emph{all blobs it contains must be included} in the block (self-evidently).
Consider, then, what would happen in the case that a transaction with a large number of blobs (e.g., 6) is submitted (showcasing a large demand for blobs from the submitter) and at the same time, a smaller but \emph{sufficiently higher-paying} transaction is already in the mempool (e.g., this transaction has high priority for only including 1 blob). Since the Ethereum block has a six-blob limit, none of the blobs from the larger transaction would be included, and only one blob would make it into the block; this is clearly and grossly inefficient.

A straightforward solution would be for the submitter to instead submit six transactions with only one blob requested in each (or any other subdivision they might deem appropriate).
However, this is an incomplete solution in that there is at least two issues with it: first, it would cause mispricing in the execution market: now, additional EIP-1559 gas would \emph{have} to be consumed; this would therefore be suboptimal.
Second, suppose for some reason\footnote{There is a natural motivation for providers wanting such adjusted utility functions for blob inclusions in more intricate settings that include, e.g., coordination of multiple rollups.} the submitter had super-additive value for simultaneous inclusion of multiple of their blobs at the same block; this naive solution would \emph{not allow} the submitter to provide such a super-additive premium for including more than one blob of theirs.
The reason for this is that, among a number of blob transactions, a block builder or validator may include their favorite subset of these.\footnote{Such a more "global" optimization on part of the block builder might make it optimal for them, in such a secnario of the naive solution, not to include blob transactions in the intended sequence by the submitting rollup or otherwise blob submitter.}

%% file: discussion.tex
\section{Improving the Blob Market}
The blob market is still in its early days and we have not consistently observed blob demand at the target. Still, type-3 transaction senders collectively pay a significant fee -- especially during congested periods such as the Blobscriptions craze and the LayerZero airdrop. However, most fees paid by type-3 transactions are either the EIP-4844 base fee or the EIP-1559 base fee, and only a small proportion of these fees is collected by the block's fee recipient. As a result, the financial incentives to include blobs are minimal. Even more so, the incentives for infrastructure investments are small and we presume that as a result, we have observed slow infrastructure development regarding blobs. One example would be the inefficiencies in blob packing we observe. Similarly, Geth (i.e., the biggest execution layer client) released a version with a bug in the blob mempool and it took nearly ten days to notice and fix~\parencite{szilagyi2024,szilagyi2024tweet2,szilagyi2024tweet3,szilagyi2024tweet4}. As a result of this bug, clients only had access to a small fraction of blob transactions. A similar bug for non-blob transactions persisting in the biggest execution layer client for more than ten days appears almost unimaginable. 

These incidents highlight the need for infrastructure investment in the blob market, for which the financial incentives currently do not appear to suffice. 

\subsection{Blob (Priority) Fees}
Some of the following steps could be taken to ensure that the ecosystem has the incentives to improve the blob market. 

\paragraph{Multidimensional (priority) fee market} Since the Decun hard fork, Ethereum has implemented a multidimensional gas market, separating the blob fee market from the general fee market. There are ongoing discussions to further divide non-blob gas usage into categories like computation, storage, and bandwidth~\parencite{buterin2024multidimensional}. Adding priority fees for each dimension would help scale fees appropriately during congestion. This approach addresses issues like those seen during the LayerZero airdrop, where EIP-1559 priority fees did not adjust adequately (see Appendix~\ref{app:prio}).

\paragraph{Increasing the blob target gradually} There are ongoing discussions within the Ethereum community about increasing the blob target soon. This increase is crucial for scaling the Ethereum ecosystem. However, a rapid target increase risks a mismatch, where demand takes months to catch up, resulting in extremely lower fees for type-3 transactions. A gradual, steady increase may help balance supply and demand. Nonetheless, creating an appropriate schedule for such a gradual increase presents its own challenges.

\paragraph{Speedup price discovery} During the LayerZero airdrop the base fee took more than six hours for price discovery to be achieved. This slow price discovery, in part, is a result of the minimum fee per blob gas being low and the maximum increase per block being 12.5\%. Thus, it is no surprise that price discovery took several hours when the base fee had to increase by 15 orders of magnitude. To address this issue, EIP-7762 seeks to increase the minimum blob fee per gas~\parencite{resnick2024eip7762,resnick2024tweet}. As a result of setting this minimum higher, the base fee would start higher and not have to move as much in times of congestion. Beyond that, this increase should not greatly impact blob transaction senders as they are still expected to pay less in the EIP-4844 gas market than they already do in the EIP-1559 gas market~\parencite{dataalways2024minimumblobfees}.

\subsection{Blob and Block Packing}
In addition to providing financial incentives to invest in the blob market, we must also outline the specific actions we can take.

\paragraph{Better packing algorithms} Our work shows that non-PBS and PBS blocks are often packed inefficiently. While some of this may result from builders avoiding blocks with blobs due to delayed propagation in the trusted-relay PBS system, many observed inefficiencies cannot be explained by this alone. If financial incentives in the blob market increase, better block-packing algorithms should emerge, especially for blob transactions. Importantly, block packing in a multidimensional market is complex, still, our proposed strategy shows potential improvements of up to 70\%. Since highly specialized builders currently construct most blocks through PBS, this added complexity is unlikely to pose a barrier.

\paragraph{Subset bidding} \Cref{subsec:market} makes the case for changing the market design for blob transactions. To fully resolve this structural market design issue that we point out, blobs would either need to go through a modified (more flexible) implementation of a mempool, or the transaction standard would need to be adjusted.
The fix that we delineate below is relatively straightforward in its description, even though it would require pervasive changes within most Ethereum clients: in its most general form, the standard should allow bidding for any subset of blobs in a transaction, and only a subset of those to be included on-chain (with the rest skipped in the final transaction).
Another way of implementing this change is to include transaction signatures for multiple subsets of blobs, in such a way that they are mutually exclusive so that only one can be chosen to be included on-chain at a time.

A special solution to this market design challenge has been introduced by Titan Builder, enabling senders to submit multiple permutations of blob transactions~\parencite{titanbuilderxyz2024api,titanbuilderxyz2024}. These blobs are then individually sorted to find the optimal combination before being included in a block.
This feature would attempt to alleviate the issue in \Cref{subsec:market} without changing the way type-3 transactions work (only for blobs submitted \emph{privately} through Titan Builder, when they win the block). While this improvement appears promising, we do not find signs of it being utilized to the extent it could and we conjecture that this might be related to the absence of congested periods since the feature's launch; we refer the interested reader to Appendix~\ref{app:private} for more details.

\section{Conclusion}

In conclusion,
while the innovation of blob transactions through Ethereum's EIP-4844 offers promising advantages, including cost reductions for L2 data settlement, our study reveals several inefficiencies.
Such inefficiencies in block packing and economic incentives seem to have emerged as issues expected to be critical in a world where the temporary storage of blobs is dominant for Ethereum. We documented periods of up to 70\% fee loss due to suboptimal blob packing, highlighting the need for better utilization of blob space which would result in both higher builder profits and less blob inclusion delay. The lack of significant financial incentives for builders to prioritize blob transactions likely contributes to these inefficiencies, as most fees associated with blob transactions are burnt, leaving builders with minimal rewards.

This work provides valuable insights into the early days of blobs on Ethereum, shedding light on both the potential and challenges of this scaling solution. As Ethereum continues to evolve, we hope for these findings to inform future network upgrades, contributing to more efficient and scalable decentralized systems.

%% file: appendix.tex
\section{Private Blob Transactions}
\label{app:private}

\begin{figure}[h]
\centering

\begin{minipage}[t]{.48\linewidth}
    \includegraphics[scale=1.2]{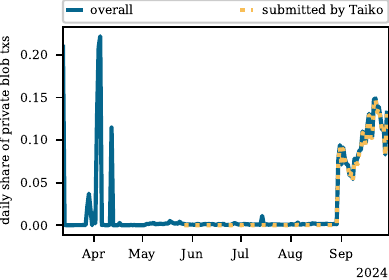}
    \caption{Proportion of privately submitted blob transactions over time. We further highlight those blob transactions privately submitted by Taiko. Notice that Taiko is responsible for nearly all privately submitted blob transactions from September onwards.}
    \label{fig:private}
\end{minipage}
\hfill
\begin{minipage}[t]{.48\linewidth}
     \includegraphics[scale=1.2]{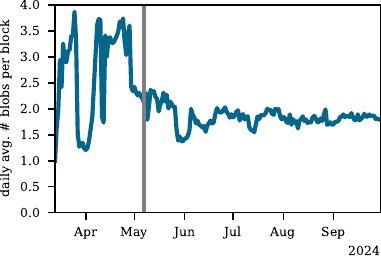}
    \caption{The daily average number of blobs per block for Titan Builder. Titan Builder implemented a new RPC call allowing the submission of multiple permutations for blob transactions on 5 July 2024 -- marked by the vertical grey line.  }
    \label{fig:titan}
\end{minipage}
\end{figure}

We use mempool data from the Mempool Dumpster project~\parencite{flashbots2024mempool} to assess whether a transaction was submitted privately or not. In Figure~\ref{fig:private}, we plot the daily share of privately submitted blob transactions. Initially, around April 2024, we observe up to 20\% of blob transactions being submitted privately. We could not identify the reason for this but presume that it is either blob transaction senders testing out various ways of submitting their transactions or the mempool loggers not being set up properly for the new transaction type yet. From May onwards we observe almost no privately submitted transactions until the start of September 2024. From then on the share or private blob transactions rose to 10\% to 15\%. The reason is that Taiko started to submit its blob transactions privately. As we highlight with the yellow dashed light essentially all private transactions from September onwards were submitted by Taiko.  

A further factor that could impact the share of privately submitted blob transactions is Titan Builder's special RPC call. The call allows subset bidding, offering more flexibility and promising higher efficiency during congested periods. Titan Builder implemented the call on 5 July 2024. Transactions submitted in this fashion would likely not hit the mempool as the additional flexibility gains would then be lost. 

In Figure~\ref{fig:titan}, we plot the daily average number of blobs per block built by Titan Builder and further indicate with the vertical grey line the data on which the RPC call was deployed. However, we did not notice an increase in the number of blobs included by Titan Builder in response to the call being implemented. Furthermore, there is also no increase in the share of private blob transactions around that time (see Figure~\ref{fig:private}), and the only rise in private transactions comes from Taiko who would not benefit from the implemented call. Taiko only submits a single blob per transaction (see Figure~\ref{fig:numBlobs}). Thus, the call does not seem to be utilized yet but this could be due to there not being any congested periods since it was implemented. 

\section{Priority Fees of Blob Transactions}\label{app:prio}

\begin{figure}[t!]
    \centering
    \begin{subfigure}[t]{0.48\columnwidth}
        \includegraphics[scale=1.2]{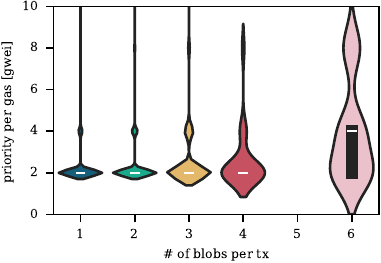}
    \caption{Blast}
    \label{fig:Blast_prio}
    \end{subfigure}\hfill
    \begin{subfigure}[t]{0.48\columnwidth}
        \includegraphics[scale=1.2]{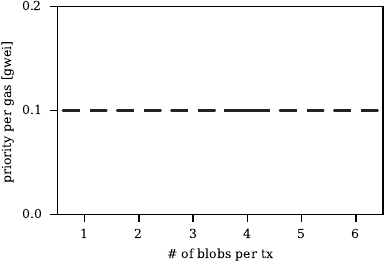}
    \caption{Starknet}
    \label{fig:starknet_prio}
    \end{subfigure}    
    \begin{subfigure}[t]{0.48\columnwidth}
        \includegraphics[scale=1.2]{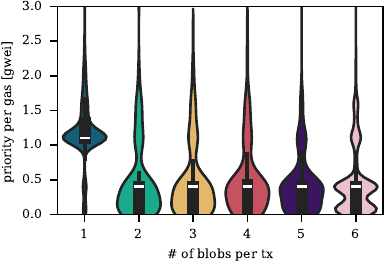}
    \caption{Linea}
    \label{fig:Linea_prio}
    \end{subfigure}\hfill
    \begin{subfigure}[t]{0.48\columnwidth}
        \includegraphics[scale=1.2]{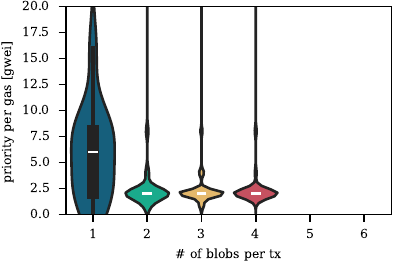}
    \caption{zkSync}
    \label{fig:zksync_prio}
    \end{subfigure}    
    \caption{Distribution of priority fee per gas for blob transactions depending on the number of blobs per transaction for Blast (see Figure~\ref{fig:Blast_prio}), Starknet (see Figure~\ref{fig:starknet_prio}), Linea (see Figure~\ref{fig:Linea_prio}), and zkSync (see Figure~\ref{fig:zksync_prio}). }\label{fig:prio_dist}
\end{figure}

In the following, we dive further into the EIP-1559 priority fees of blob transactions. Recall, that priority fees do not scale with the number of blobs but with the EIP-1559 gas usage. We place a particular focus on the senders that submit blob transactions with different numbers of blobs simultaneously for extended periods: Blast, Starknet, Linea, and zkSync (see Figure~\ref{fig:numBlobs}). Given that these rollups submit different numbers of blobs per transaction at the same time, we can compare the priority fees they indicate when focusing on those periods in particular.

In Figure~\ref{fig:prio_dist}, we plot the distribution of the priority fee per gas for blob transactions from these senders and restrict the analysis to periods where the respective sender was submitting differing numbers of blobs per transaction. We find that the priority fee per gas is not strongly positively correlated with the number of blobs for any of the rollups we look at. Instead, we the correlation to be either weakly positive for Blast, no correlation for Starknet, and weakly negative for Linea and zkSync.

Nonetheless, it is important to highlight that gas usage---especially for ZK rollups---can correlate with the number of blobs. As a result, the priority fee might also scale with the number of blobs and the priority fee per gas would not necessarily be the correct metric to consider. Thus, we calculate total the priority fee for each blob transaction. 

\begin{figure}[t]
    \centering
    \includegraphics[scale=1.2]{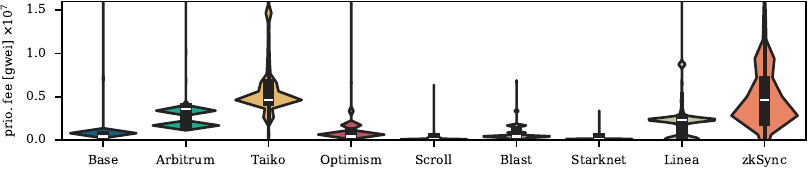}
    \caption{Distribution of total priority fee for blob transactions depending on the transaction sender.}
    \label{fig:total_prio_per_rollup}
\end{figure}

We first investigate the distribution of this measure for each of the biggest blob users in Figure~\ref{fig:total_prio_per_rollup}. Here we notice that Taiko pays the highest priority fee per transaction on average, with zkSync, Linea, and Arbitrum also paying more than average in this measure. The remaining rollups have a quite small total priority fee. For some (i.e., Base, Optimism, and Blast), this is due to them being optimistic rollups and using very little gas (see Figure~\ref{fig:gas_used}), while for Scroll and Starknet this is due to them paying little priority fees per gas in general (see Figure~\ref{fig:prio}).

\begin{figure}[t]
    \centering
    \begin{subfigure}[t]{0.48\columnwidth}
        \includegraphics[scale=1.2]{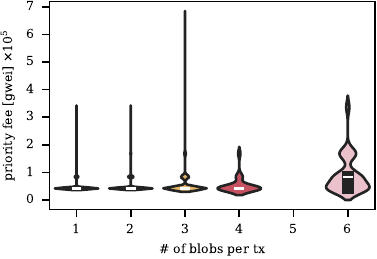}
    \caption{Blast}
    \label{fig:Blast_prio_per_blob}
    \end{subfigure}\hfill
    \begin{subfigure}[t]{0.48\columnwidth}
        \includegraphics[scale=1.2]{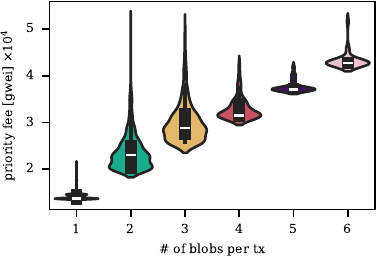}
    \caption{Starknet}
    \label{fig:starknet_prio_per_blob}
    \end{subfigure}    
    \begin{subfigure}[t]{0.48\columnwidth}
        \includegraphics[scale=1.2]{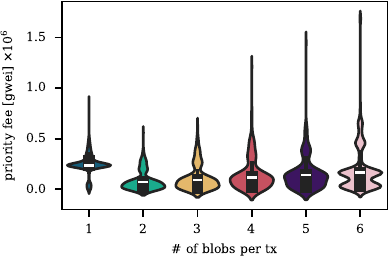}
    \caption{Linea}
    \label{fig:Linea_prio_per_blob}
    \end{subfigure}\hfill
    \begin{subfigure}[t]{0.48\columnwidth}
        \includegraphics[scale=1.2]{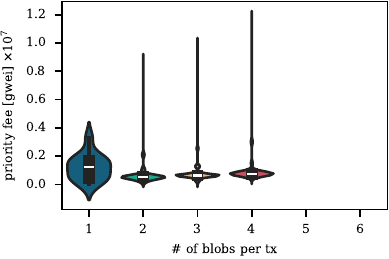}
    \caption{zkSync}
    \label{fig:zksync_prio_per_blob}
    \end{subfigure}    
    \caption{Distribution of total priority fee for blob transactions depending on the number of blobs per transaction for Blast (see Figure~\ref{fig:Blast_prio_per_blob}), Starknet (see Figure~\ref{fig:starknet_prio_per_blob}), Linea (see Figure~\ref{fig:Linea_prio_per_blob}), and zkSync (see Figure~\ref{fig:zksync_prio_per_blob}). }\label{fig:prio_per_num_blobs}
\end{figure}

We also investigate the total priority fee for the four rollups submitting blob transactions with various numbers of blobs simultaneously for an extended period (see Figure~\ref{fig:prio_per_num_blobs}).
If rollups set the priority fee per gas in such a way that correlates positively with the number of blobs they would like to include, then we should expect to see this positive correlation between the total priority fee and the number of blobs in Figure~\ref{fig:prio_per_num_blobs}. However, we observe that this is not generally the case. Instead, for three of the rollups (Blast, Linea, and zkSync), we observe a generally insignificant (mildly negative or positive) correlation between the total priority fee and the number of blobs. We do observe a strong positive correlation of 0.92 for Starknet, but importantly correlation between the gas usage and the number of blobs is also exactly 0.92. Thus, the EIP-1559 gas usage is correlated with the blob gas usage for the rollup, and even without explicitly setting the priority fee in a way that scales with the number of blobs this would occur implicitly. Together this analysis indicates that the rollups do not appear to set the priority fee in a way that scales with the number of blobs they include in a transaction.

Such behavior is predicted especially in times when the blob market is congested~\parencite{buterin2024multidimensional,dataalways2024minimumblobfees}. The intuition is that when the blob market is congested and the blob base fee is still in price discovery, the blob transaction senders compete in a first-price auction. Given that blobs are the congested resource in this example, they would adjust the priority fee in a way that it would scale with the blob gas usage (i.e., the number of blobs) and then compete on the priority fee per blob measure. 

\begin{figure}[H]%
\centering
    \includegraphics[scale=1.2]{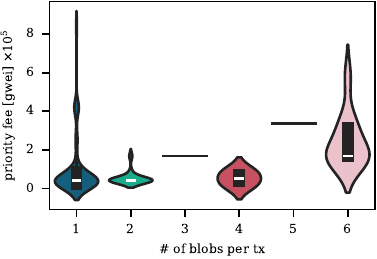}
  \caption{Distribution of total priority fee for blob transactions depending on the number of blobs per transaction for zkSync during the LayerZero airdrop.}\label{fig:zksync_prio_per_blob_airdrop}
\end{figure}

The previous analysis, however, did not focus on congested periods. Thus, to analyze whether we observe this predicted behavior from the rollups we focus on zkSync during the congested period surrounding the LayerZero airdrop. zkSync was the only rollup already submitting blob transactions with different numbers of blobs during that period. We plot the distribution of total priority fee for blob transactions depending on the number of blobs per transaction in Figure~\ref{fig:zksync_prio_per_blob_airdrop}.
We find that the priority fee per blob seems more or less constant independent of the number of blobs, there is only a mild positive correlation of 0.14 which is not much higher than zkSync's correlation between gas usage and the number of blobs at 0.10. Thus, zkSync does not seem to be adjusting the priority fee for the number of blobs. As a result, our analysis is (partially) inconclusive in regard to how the rollups set their priority fees during congested periods but find little to no signs of them doing so.